\documentclass[aps,preprint]{revtex4}
\usepackage{amsfonts}
\usepackage{amsmath}
\usepackage{amssymb}
\usepackage{graphicx}%
\providecommand{\U}[1]{\protect\rule{.1in}{.1in}}

\begin{document}
\preprint{HEP/123-qed}
\title[A STD description of Quantum many-body wavefunction]{\textbf{Symmetric Tensor Decomposition Description of Fermionic Many-Body
Wavefunctions}}
\author{Wataru Uemura and Osamu Sugino}
\affiliation{The Institute for Solid State Physics, The University of Tokyo, 5-1-5
Kashiwanoha, Kashiwa, Chiba, 277-8581, Japan}
\pacs{PACS number}

\begin{abstract}
The configuration interaction (CI) is a versatile wavefunction theory for
interacting fermions but it involves an extremely long CI series. Using a symmetric
tensor decomposition (STD) method, we convert the CI series into a compact and
numerically tractable form. The converted series encompasses the Hartree-Fock
state in the first term and rapidly converges to the full-CI state, as
numerically tested using small molecules. Provided that the length of the STD-CI
series grows only moderately with the increasing complexity of the system, the new
method will serve as one of the alternative variational methods to achieve full-CI with enhanced practicability.

\end{abstract}
\maketitle


An accurate description of the ground-state wavefunction of an interacting
Fermion system is one of the central goals of modern science. The most straightforward and versatile approach to 
describing this wavefunction is the configuration
interaction (CI),
but its numerical application is greatly limited by the fact that the full-CI
series consists of $_{M}C_{N}$ Slater determinants (SDs) when describing an
$N$-electron system using $M$ basis functions. To truncate this extremely
long CI series without compromising on chemical accuracy, many methods have
been developed, such as the multi-reference CI, which uses a part of the SDs derived
from a few of the most important ones, or the complete active space (CAS) CI which
uses all SDs generated from a selected set of orbitals \cite{CI}. Even so, the
application has been hampered by the slow convergency of the CI series.

In this context, the many-body perturbation approaches to treat all SDs have
attracted attention; these approaches include the coupled cluster (CC) theory \cite{CC1}%
\cite{CC2} which is used to represent the wavefunction in terms of an SD (or a few SDs)
applied with the exponential of an excitation operator. The CC theory has proven
accurate for a number of molecules, although it occasionally provides qualitatively
incorrect potential surfaces \cite{Nakata}. The density matrix renormalization
group (DMRG) method \cite{DMRG} has also attracted attention as a variational
method within the space of the matrix product state \cite{MatrixProductState}.
It has been extensively applied to correlated electron systems 
\cite{ApplicationOfDMRG1}\cite{ApplicationOfDMRG2}; however, this method was
originally formulated only for one-dimensional systems and its extension to
three-dimensional systems is not very straightforward.

Recent tensor analyses have shown that, despite the large number, the CI
coefficients may be described by a tractable number of variational parameters.
For example, the full-CI results of some molecules were accurately
reproduced by the complete-graph tensor network (CGTN) state containing $\sim M^{2}$ variational parameters \cite{CGTN}\cite{CGTNII}. Tensor
decomposition (TD) \cite{PARAFAC2} methods such as the Tucker decomposition
\cite{Tucker} and the canonical decomposition (CANDECOMP)/parallel factor
decomposition (PARAFAC), abbreviated as CP, \cite{PARAFAC3} \cite{PARAFAC}
have also been applied to molecules. These methods were used to analyze the double
excitation tensor $\mathcal{T}_{2}$ originating from the electron-electron
interaction \cite{MP2TUCKER}\cite{MP2CP}. The results showed that the
$\mathcal{T}_{2}$ tensor of rank 4, consisting of $\sim M^{4}$terms, can be
described by $\sim MK$ parameters, where $K$ denotes the length of the tensor
decomposition \cite{MP2CP}. The TD method was also suggested as being effective in
greatly reducing the variational parameters required for full-CI
\cite{MP2TUCKER}.

In this context, we formulate a practical scheme to perform full-CI level
calculation using a TD method. In this study, we describe the CI coefficients as a
product of a symmetric tensor and the permutation tensor, and following the CP procedure we
expand the former into $K$ symmetric Kronecker product states, which are
composed of vectors of dimension $M$. Subsequently, we calculate the second-order
density matrix consisting of $\sim K^{2}M^{4}$ elements using the
Vieta's formula \cite{Vieta} thereby performing $\sim M^{2}$ operations for each
element. This allows us to perform the total energy calculation variationally
using $\sim K^{2}M^{6}$ operations. Our test calculations for the potential
surface of simple diatomic molecules and for a Hubbard cluster model with
different parameters show that with increasing $K$,$\,$the total energy
rapidly converges to the full-CI result. This shows that our symmetric
tensor decomposition CI (STD-CI) scheme will greatly extend the applicability
of the full-CI level calculation, provided that $K$ increases only moderately with
$N$, $M$, or the complexity of the electron correlation. In the rest of this paper, we provide the details of STD-CI.


We begin by describing the CI-series representation of the many-body
wavefunction%
\begin{equation}
\Psi\left(  x_{1}\cdots x_{N}\right)  =\sum_{i_{1}\cdots i_{N}=1}^{M}%
A_{i_{1}\cdots i_{N}}\psi_{i_{1}}\left(  x_{1}\right)  \cdots\psi_{i_{N}%
}\left(  x_{N}\right)  , \label{ManyBodyWF}%
\end{equation}
where $x_{1}\cdots x_{N}$ represent the space and spin coordinates of the electrons and
$\psi_{i_{k}}$s are the orthonormal orbitals which are represented as a linear
combination of orthonormalized basis functions as
\[
\psi_{i}\left(  x\right)  =\sum_{j}U_{ij}\phi_{j}\left(  x\right)  .
\]
The antisymmetric tensor $A_{i_{1}\cdots i_{N}}$ can be described as the product
of a symmetric tensor ($S_{i_{1}\cdots i_{N}}$) of rank $N$ and dimension $M$%
\ and a product of $N\left(  N-1\right)  /2$ permutation tensors
($\epsilon_{ij}$s) of rank $2$ as%

\begin{equation}
A_{i_{1}\cdots i_{N}}=S_{i_{1}\cdots i_{N}}\epsilon_{i_{1}i_{2}}%
\epsilon_{i_{1}i_{3}}\cdots\epsilon_{i_{N-1}i_{N}}\text{.} \label{SOPD1}%
\end{equation}
Next, $S_{i_{1}\cdots i_{N}}$\ is decomposed into a minimal linear combination
of symmetric Kronecker product states using vectors of dimension $M$,
$c_{i_{k}}^{1},\cdots,c_{i_{k}}^{K}$, as%
\begin{equation}
S_{i_{1}\cdots i_{N}}=\sum_{j=1}^{K}\lambda_{j}c_{i_{1}}^{j}\cdots c_{i_{N}%
}^{j}. \label{SOPD2}%
\end{equation}
This symmetric tensor decomposition (STD) is a symmetric version of CP, which
is also a special case of the symmetric Tucker decomposition%
\begin{equation}
S_{i_{1}\cdots i_{N}}=\sum_{j_{1}\cdots j_{N}}s_{j_{1}\cdots j_{N}}u_{i_{1}%
}^{j_{1}}\cdots u_{i_{N}}^{j_{N}} \label{Tucker}%
\end{equation}
in that the transformed tensor $s_{j_{1}j_{2}\cdots j_{N}}$ is the superdiagonal
$\lambda_{j}$ in CP. The total energy is optimized by varying the vectors
$c_{i}^{j}$ and the unitary matrix $U_{ij}$, so that no approximation is made
in our STD-CI method apart from the truncation of the series  at $K$.
It is noteworthy that each term in the STD series contains all the SDs generated from the
orbitals $\psi_{i}$, although the degrees of freedom are only $MK+M\left(
M+1\right)  /2$ as a whole, thereby indicating that we are treating the entangled
states and that the degree of entanglement is reduced with increasing $K$. It can be shown that the 
Hartree-Fock (HF) approximation corresponds to taking $K=1$ and $c_{N+1}%
^{1}=\cdots=c_{M}^{1}=0$; therefore, the approximation with $K=1$ is already a
natural extension of the HF approximation. When treating a weakly correlated system, an HF-like
solution is obtained and on the other hand, when treating a strongly correlated system, an orbital-ordered solution
is obtained, provided that a sufficiently
large value of
$K$ is considered. In this manner, we can bridge the HF solution with the fully correlated
state by increasing the value of $K$. The STD-CI will be exact when
$K = _MC_N$.


Our numerical procedure begins by constructing the second-order density matrix
(DM), which has the form%
\[
\gamma_{2}\left(  x_{1}x_{2},x_{3}x_{4}\right)  =\sum_{i_{1}i_{2}i_{3}i_{4}%
=1}^{M}\Gamma_{i_{1}i_{2}i_{3}i_{4}}\psi_{i_{1}}^{\ast}\left(  x_{1}\right)
\psi_{i_{2}}^{\ast}\left(  x_{2}\right)  \psi_{i_{3}}\left(  x_{3}\right)
\psi_{i_{4}}\left(  x_{4}\right)  .
\]
Using (\ref{SOPD1}) and (\ref{SOPD2}), the DM coefficient can be rewritten as%
\[
\Gamma_{i_{1}i_{2}i_{3}i_{4}}=\sum_{ij=1}^{K}\lambda_{i}\lambda_{j}c_{i_{1}%
}^{i\ast}c_{i_{2}}^{i\ast}c_{i_{3}}^{j}c_{i_{4}}^{j}\epsilon_{i_{1}i_{2}%
}\epsilon_{i_{3}i_{4}}I_{i_{1}i_{2}i_{3}i_{4}}^{ij}%
\]
where $I$ for each set of indices $\left\{  iji_{1}i_{2}i_{3}i_{4}\right\}  $
is expressed, using $a_{k_{l}}\equiv c_{k_{l}}^{i\ast}c_{k_{l}}^{j}\epsilon
_{i_{1}k_{l}}\epsilon_{i_{2}k_{l}}\epsilon_{i_{3}k_{l}}\epsilon_{i_{4}k_{l}}$,
as
\begin{equation}
I=\sum_{k_{3}\cdots k_{N}=1}^{M}a_{k_{3}}\cdots a_{k_{N}}\left(
\epsilon_{k_{3}\cdots k_{N}}\right)  ^{2}. \label{RDstep1}%
\end{equation}
Based on the fact that the permutation tensor squared is equal to $1$ when all the
indices are different and $0$ otherwise, it can be shown using
Vieta's formula that the value of $I$ is equal to the $M-\left(  N-2\right)
$-th order coefficients of the polynomial $\left(  N-2\right)  !f_{M}\left(  t\right)  $
with $f_{M}\left(  t\right)  \equiv\left(  t+a_{1}\right)  \cdots\left(
t+a_{M}\right)  $ [19]. 
The coefficient can be easily obtained by using a list manipulation, where the coefficients of $f_{p}\left(
t\right)  $ with $0\leq p\leq M$ are described by a row-vector of dimension $p$ as $\boldsymbol{f}_{p}=\left(  f_{p,0},f_{p,1},\cdots,f_{p,p-1}\right)  $ and are applied with the iterative equation, $\boldsymbol{f}_{p}=a_{p}\left(
\boldsymbol{f}_{p-1},0\right)  +\left(  0,\boldsymbol{f}_{p-1}\right)  $, considering $\boldsymbol{f}_{0}=1$. $\propto M^{2}$ operations are required to obtain the coefficient of $f_{M}(t)$. Therefore, the total number of operations needed to obtain all
$I$'s is $\propto K^{2}M^{6}$. In practical coding, one may use the fact that $a_{i_k}=0$ when $i_k$ is equal to one of the four indices $\{ i_1 i_2 i_3 i_4 \}$ to achieve further efficiency.

Subsequently the parameters $c_{i_{k}}^{j}$, $\lambda^{j}$, and $U_{ij}$ are varied to
minimize the total energy $E_{tot}=\sum h_{i_{1}i_{2}i_{3}i_{4}}\Gamma
_{i_{3}i_{4}i_{1}i_{2}}/\sum\Gamma_{i_{1}i_{2}i_{1}i_{2}}$ with%
\begin{eqnarray}
h_{i_{1}i_{2}i_{3}i_{4}}&=&\int dx_{1}dx_{2}\psi_{i_{1}}^{\ast}\left(
x_{1}\right)  \psi_{i_{2}}^{\ast}\left(  x_{2}\right)  \left[  N\left(
-\frac{1}{2}\boldsymbol{\nabla}_{1}^{2}+v_{ext}\left(  \boldsymbol{r}%
_{1}\right)  \right)  +\frac{N\left(  N+1\right)  }{2}\frac{1}{\left\vert
\boldsymbol{r}_{1}-\boldsymbol{r}_{2}\right\vert }\right]  \nonumber \\ 
&\times &\psi _{i_{3}}\left(
x_{1}\right)  \psi_{i_{4}}\left(  x_{2}\right)  \text{,}
\end{eqnarray}
where $v_{ext}$ denotes the external potential. 
In the variation, we require derivatives of $I_{i_1 i_2 i_3 i_4}^{ij}$ with respect to $a_{i_k}$ for those $i_k$ not in $\{i_1 i_2 i_3 i_4 \}$. To obtain the derivatives, we need to differentiate $f_M (t)$ by $a_{i_k}$ and obtain its $M-(N-1)$-th coefficient. When this is done simply using the list manipulation, $\propto K^{2}M^{7}$ operations are required for each $i_k$; however, the number of operations can be reduced when using $f_{M}\left(
t\right)  /\left(  t+a_{i_{k}}\right)  $ for the differentiation. When the series $(t+a_{i_{k}})^{-1}=\sum_{m=0}^{\infty}a_{i_{k}}^{-m-1}\left(  -t\right)
^{m}$ is multiplied with $f_{M}\left(
t\right)  $, the $M-(N-1)$-th coefficient can be obtained as
\[
-\sum_{s=0}^{M-\left(  N-1\right)  }\left(  -a_{i_{k}}\right)  ^{s-1-M+\left(
N-1\right)  }f_{M,s}\text{,}%
\]
thereby requiring $\propto M$ operations for each $k$. Therefore, $\sim K^2 M^6$ operations are required to obtain all the derivatives of $I_{i_1 i_2 i_3 i_4}^{ij}$.
By applying the same technique to the expression $f\left(
t\right)  /\left(  t+a_{i}\right)  \left(  t+a_{j}\right)  $,$\ $the second
derivatives are similarly obtained with $\propto K^{2}M^{6}$ operations. It should be noted
that the calculation of the derivatives is the rate-determining step in our calculation.

In a manner similar to HF \cite{Brener},\cite{Fry}, STD-CI can be applied to a crystalline solid
by taking a linear combination of the atomic orbitals $\chi_{i}$ as%
\[
\phi_{ik}\left(  r\right)  =\sum_{\tau}e^{ik\cdot R_{\tau}}\chi_{i}\left(
r-R_{\tau}\right)  ,
\]
where $k$ and $R_{\tau}$ denote the reciprocal vector in the Brillouin zone and
the nuclear coordinate, respectively. Thus, $M$ should be read as the number
of $k$ values multiplied by the number of basis functions.


To assess the efficiency of STD-CI, we investigate how many terms in Eq.(\ref{SOPD2}) are required to
achieve convergence in $E_{tot}$. This investigation is carried out for simple diatomic molecules
(H$_{2}$, He$_{2}$, and LiH) and a four-site Hubbard model. Relativistic effects are neglected and only
 the spin
unpolarized state is calculated by using the same number of orbitals
with an $\alpha $ and $\beta $ spin.
In testing the
convergence, the calculated results are compared with the full CI calculation
performed using the same basis functions. In our calculations, the
Newton-Raphson method is used to variationally determine the parameters.

\begin{figure}
[ptb]
\begin{center}
\includegraphics[
height=2.0in,
width=3.0in
]%
{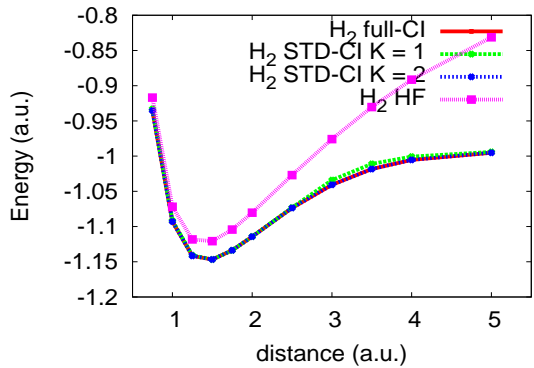}%
\caption{Potential curve of H$_{2}$ molecules for full-CI (solid line), STD-CI
with $K=1$ (broken line with cross) and $K=2$ (broken line with asterisk), and
HF (dotted line).}%
\label{fig1}%
\end{center}
\end{figure}

\begin{figure}
[ptb]
\begin{center}
\includegraphics[
height=2.0in,
width=3.0in
]%
{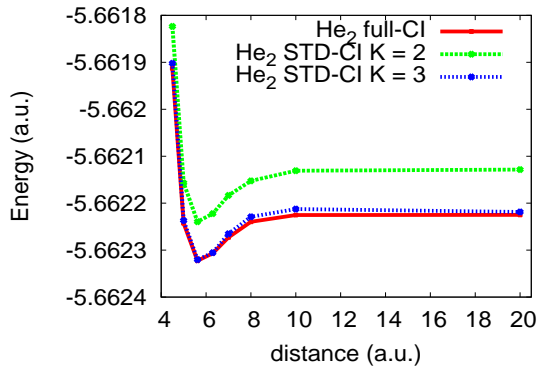}%
\caption{Potential curve of He$_{2}$ molecules for full-CI (solid line) and
STD-CI with $K=2$ (broken line with cross) and $K=3$ (broken line with
asterisk).}%
\label{fig2}%
\end{center}
\end{figure}

\begin{figure}
[ptb]
\begin{center}
\includegraphics[
height=2.0in,
width=3.0in
]%
{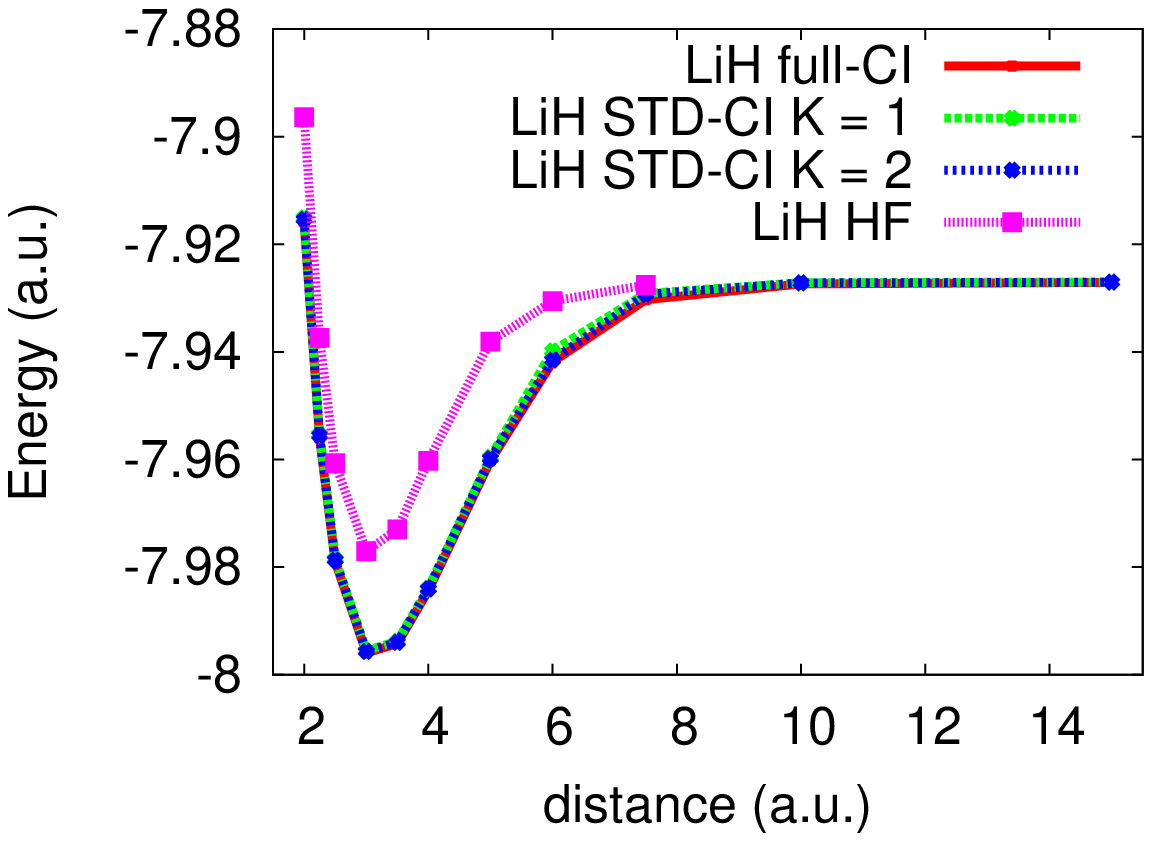}%
\caption{Potential curve of LiH molecules for full-CI (solid line), STD-CI
with $K=1$ (broken line with cross) and $K=2$ (broken line with asterisk), and
HF (dotted line).}%
\label{fig3}%
\end{center}
\end{figure}

\begin{figure}
[ptb]
\begin{center}
\includegraphics[
height=2.0in,
width=3.0in
]%
{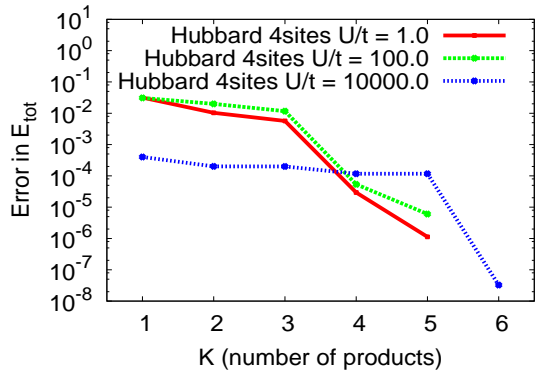}%
\caption{Error in E$_{tot}$ for the four-site Hubbard model for various
parameters $U/t=1$ (solid line), $U/t=100$ (dotted line with cross), and
$U/t=10000$ (dotted line with asterisk).}%
\label{fig4}%
\end{center}
\end{figure}

H$_{2}$ is the simplest molecule where the molecular orbital picture, valid
near the equilibrium bond length, is switched to the Heitler-London picture,
as the interatomic distance increases to infinity. The calculation with the STO-3G
basis set shows that $K=1$ reproduces the full CI potential curve within an error of 0.01
Ha error, while the error is less than 0.01 mHa when $K=2$ (Fig.\ \ref{fig1}).
The molecule He$_{2}$ is weakly bound the dispersion forces and the test is more
stringent in this case. The calculation with the 6-311G basis set shows that $K=3$ is
sufficient to reproduce the full CI result within an error of 0.01 mHa, while $K=2$
is already sufficient to obtain the binding energy within the same accuracy
although the absolute value of $E_{tot}$ is always larger by 0.1 mHa
(Fig.\ \ref{fig2}). The binding energy is about three times larger than the
accurate quantum chemical calculation\cite{HeCalc} and the experimental results
\cite{KeineKathofer}, which is presumably due to the insufficient number of basis
functions; obtaining an accurate value
of the binding energy
is beyond the scope of our comparative
study, and this must be the consideration of future studies. LiH is a typical hetero-nuclear
diatomic molecule. The 4-31G calculation for this case shows that $K=1$ nearly sufficiently
reproduces the full CI result while the HF calculation significantly underestimates the
binding energy (Fig.\ \ref{fig3}). The final test is 
the application of our idea to
the four-site Hubbard
model in the tetrahedron structure under the half-filled condition. As the
Hubbard $U$ over the transfer $t$ increases, larger $K$
values are
required; however, $K=6$
is found sufficient even in the large $U/t$ limit (Fig.\ \ref{fig4}).

\begin{figure}
[ptb]
\begin{center}
\includegraphics[
height=2.0in,
width=3.0in
]%
{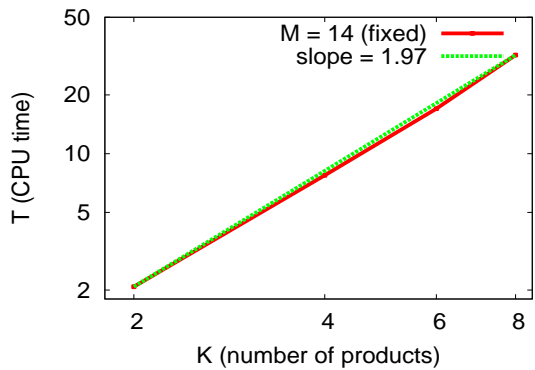}%
\caption{Total CPU time ($T$) versus $K$. The solid line
indicates result after measurement
for one iteration step and broke line
indicates that
after fitting to $T=aK^{b}$. $b$ is obtained as
1.97. }%
\label{fig5}%
\end{center}
\end{figure}

\begin{figure}
[ptb]
\begin{center}
\includegraphics[
height=2.0in,
width=3.0in
]%
{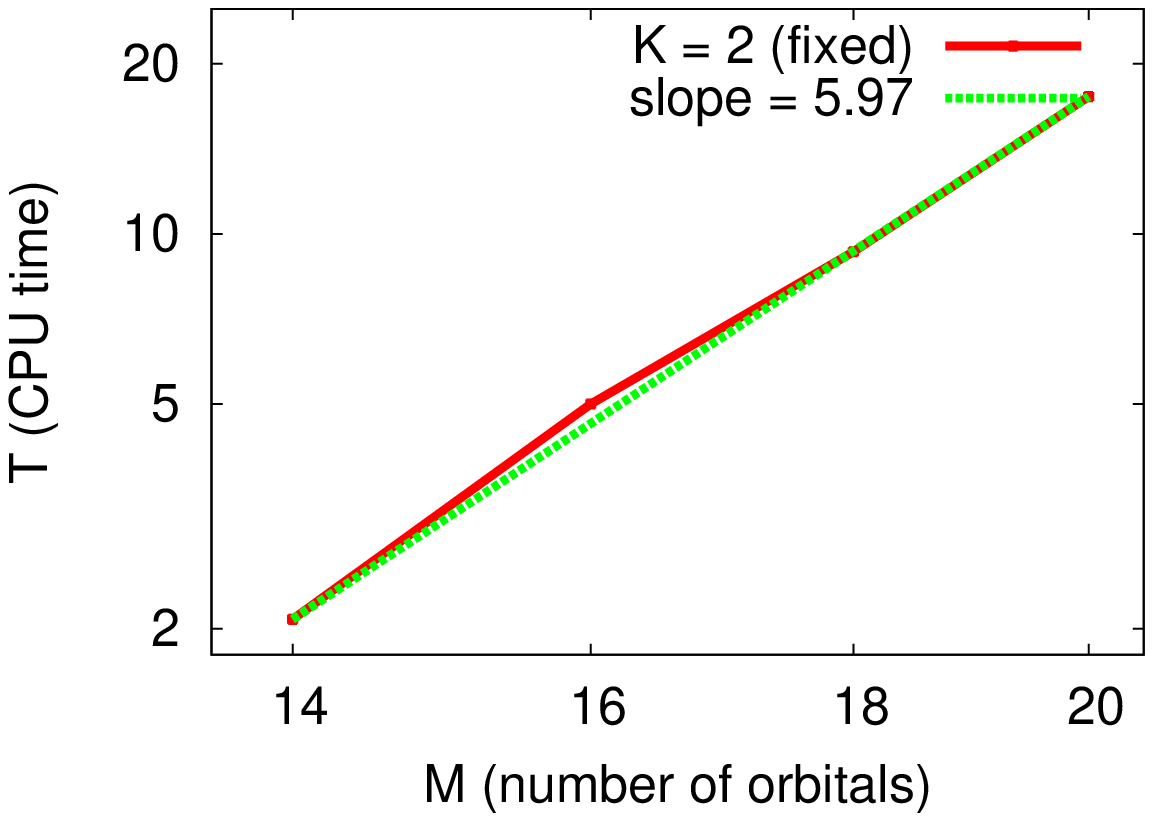}%
\caption{Total CPU time ($T$) versus $M$. The solid line
indicates result 
after measurement for one iteration step and the broken line indicates that after fitting to $T=aM^{b}$. $b$ is 5.97 after the
fitting. }%
\label{fig6}%
\end{center}
\end{figure}

The computational time theoretically scales as $K^{2}M^{6}$. We tested the time scaling with
our numerical code to find that CPU time indeed scales as $K^{1.97}M^{5.97}$
on average (Figs.\ \ref{fig5},\ref{fig6}). Because the operations involved in the calculation can be performed
independently, this method is suitable for massively parallel computers.%



We formulated a practical scheme to perform the full-CI level calculation
using the STD method. In the STD-CI method,
we expand the CI coefficients as the product of a symmetric tensor and the
permutation tensor, and we further expand the symmetric tensor into Kronecker product states
composed of vectors of dimension $M$. By varying the vectors and the unitary
transformation matrix $U_{ij}$, the total energy is minimized using the
second-order density matrix technique. The STD-CI method, 
which involves taking the
length of the
series $K$ as the only input parameter, allows us to perform a full-CI level
calculation rigorously using $KM+M\left(  M+1\right)  /2$ variational
parameters and $\sim K^{2}M^{6}$ operations. By applying the scheme to the
potential curve of small diatomic molecules such as H$_{2}$, He$_{2}$, and LiH, and
the four-site Hubbard model for various parameters, we found that a very small
$K$ value is required to reproduce the full-CI results within milli-Hartree
accuracy. If $K$ increases moderately with $N$, $M$, or the degree of
correlation, the scheme will greatly extend the applicability of the full-CI level
calculation. We believe that application of the scheme to a crystalline
solid and the use of the scheme as a building block of the fragment molecular
orbital (FMO) scheme can be of high significance\cite{FMO}.

\textbf{Acknowledgement}

The authors thank Prof. Y. Mochizuki (Rikkyo Univ.) and M. Nakata (RIKEN) for
their valuable discussions.

\end{document}